\documentclass[aps,prb,reprint,preprintnumbers,amsmath,amssymb,floatfix]{revtex4-1}

\usepackage{graphicx}
\usepackage{dcolumn}
\usepackage{bm}
\usepackage{color}
\usepackage{tabularx}
\usepackage{amsmath,amsfonts,amsthm,amssymb}
\usepackage{appendix}
\usepackage[bookmarksnumbered,pdfpagelabels=true,plainpages=false,colorlinks=true,linkcolor=blue,citecolor=blue,urlcolor=blue]{hyperref}
\usepackage[english]{babel}

\begin{document}


\title{Magnon valley Hall effect in CrI$_3$-based vdW
  heterostructures}

\author{R. Hidalgo-Sacoto$^{1}$}
\author{R. I. Gonzalez$^{2}$}
\author{E. E. Vogel$^{3,4}$}
\author{S. Allende$^{4,5}$}
\author{Jos\'e D. Mella$^{4,6}$}
\author{C. Cardenas$^{4,6}$}
\author{Roberto E. Troncoso$^{7}$}
\author{F. Munoz$^{4,6}$}
\email{fvmunoz@gmail.com}

\affiliation{$^{1}$School of Physical Sciences and Nanotechnology, Yachay Tech, Urququ\'i, Ecuador}
\affiliation{$^{2}$Centro de Nanotecnolog\'ia Aplicada, Facultad de Ciencias, Universidad Mayor, Santiago, Chile}
\affiliation{$^{3}$Departamento de Ciencias F\'isicas, Universidad de
  La Frontera, Temuco, Chile, Chile}
\affiliation{$^{4}$Center for the Development of Nanoscience and
  Nanotechnology, CEDENNA, Santiago, Chile}
\affiliation{$^{5}$Departamento de F\'isica, Universidad de Santiago
  de Chile, Santiago, Chile}
\affiliation{$^{6}$Departamento de F\'isica, Facultad de Ciencias,
  Universidad de Chile, Santiago, Chile}
\affiliation{$^{7}$Center for Quantum Spintronics, Department of
  Physics, Norwegian University of Science and Technology, NO-7491
  Trondheim, Norway}

\date{\today}

\begin{abstract}
Magnonic excitations in the two-dimensional (2D) van der Waals (vdW)
ferromagnet CrI$_3$ are studied. We find that bulk magnons exhibit a
non-trivial topological band structure without the need for
Dzyaloshinskii-Moriya (DM) interaction. This is shown in vdW
heterostructures, consisting of single-layer CrI$_3$ on top of
different 2D materials as MoTe$_2$, HfS$_2$ and WSe$_2$. We find
numerically that the proposed substrates modify substantially the
out-of-plane magnetic anisotropy on each sublattice of the CrI$_3$
subsystem. The induced staggered anisotropy, combined with a proper
band inversion, leads to the opening of a topological gap of the
magnon spectrum. Since the gap is opened non-symmetrically at the  ${\bf K}^+$ and ${\bf K}^-$ points of the Brillouin zone, an imbalance in the magnon
population between these two valleys can be created under a driving
force. This phenomenon is in close analogy to the so-called valley
Hall effect (VHE), and thus termed as magnon valley Hall effect
(MVHE). In linear response to a temperature gradient we quantify this
effect by the evaluation of the temperature-dependence of
the magnon thermal Hall effect. These findings open a different avenue
by adding the valley degrees of freedom besides the spin, in the study
of magnons.
\end{abstract}

\maketitle

\section{\label{sec:intro} Introduction}
Magnons, the low-energy spin excitations of magnets, occupy a central
place in the field of spintronics \cite{Chumak2015}. Since magnons
carry spin angular momentum and do posses electric charge, the
understanding and control of their transport properties are of
paramount importance \cite{Kruglyak2010,Serga2010}. From a practical
perspective, the lack of charge transport, implying the absence of lost
of energy in the form of heat via Joule heating is certainly
attractive.

Research on topological matter inspired a plethora of theoretical
predictions of topological magnons systems during the last years. Among the first proposals showed that an engineered magnonic
crystals develop topological bulk magnon bands and hence host chiral
edge states \cite{Shindou2013}. Other alternative routes exploit
mechanisms based on emergent gauge fields induced by magnetic
textures, e.g., Skyrmion crystals
\cite{NagaosaNature2013,FertNature2018}. These magnetic phases provide
a natural crystalline enviorement and shown that magnons inherit a
topologically nontrivial band structure
\cite{SchwarzeNature2015,Roldan2016,Garst2017}. Interestingly,
topological features are also present in certain lattice geometries
like for instance,
honeycomb-\cite{Kim2016,Owerre2016,Cheng2016,Zyuzin2016},
Kitaev-\cite{McClarty2018,Joshi2018} or Kagom\'e-lattice
\cite{Chisnell2015,Seshadri2018,Owerre2018b,Malz2019} spin systems. In
most of them the Dzyaloshinskii-Moriya (DMI) interaction is a key
element \cite{Kim2016,Zhang2013}, since it plays an analogous role to
the spin-orbit coupling(SOC) in the Kane-Mele model
\cite{Kane2005}. However, theoretical studies have shown that magnonic
bulk bands carry nontrivial Chern numbers, under the presence of
nearest-neighbor pseudodipolar interaction
\cite{Wang2017,Wang2018a,Wang2018b} and without the need for DMI.

In this work, we show that van der Waals (vdW) heterostructures,
consisting of CrI$_3$ on top of different 2D materials with a
hexagonal lattice open a topological gap in the magnon spectrum of the
CrI$_3$ subsystem.  Unlike previous schemes\cite{Kim2016,Zhang2013}, based on the
Dzyaloshinskii-Moriya coupling, our approach closely resembles  that of an
electrically-induced band gap in bilayer graphene \cite{Munoz16,McCann2013,Ju2015}. The
underlying mechanism that give arise this effect can be established by
symmetry arguments. The honeycomb lattice of CrI$_3$ has a sublattice
symmetry, with two identical but nonequivalent Cr atoms, see
Fig. \ref{fig:str}A. If we consider the heterostructure CrI$_3|$MX$_2$,
being MX$_2$ the matched hexagonal material (\textit{e.g.} a transition metal dichalcogenide), each Cr sublattice will
have a different environment and thus, the sublattice symmetry is
broken. Accordingly, the octahedrons of I atoms that wrap each Cr, are
distorted differently for each sublattice. Therefore, the
magnetocrystalline anisotropy energy will be different for each
sublattice. It is worth commenting that this is independent of
details such as the actual minimum energy configuration.

As a further step, we study the low-energy magnetic fluctuations of
the effective spin system. We focus on magnonic excitations around the
collinear ferromagnetic ground state, where the effects of the
substrate appear as a staggered on-site energy  added to the magnon
Hamiltonian. Concretely, we construct the magnonic analogue of the
valley Hall effect in graphene, breaking the inversion symmetry. In
order to discuss the experimental accessibility of the predicted
phenomena, we study magnon transport in linear response to a thermal
bias. The non-trivial Berry curvature leads to the magnon thermal Hall
effect, which is determined by the calculation of the transverse
thermal conductivity at finite temperature.

The paper is outlined as follows. In Sec. \ref{sec:methods} we
describe the numerical methods employed in this work. In
Sec. \ref{sec:model} we first determine the most energetically
favorable crystalline configuration  between the CrI$_3$
layer and each of the
proposed substrates.  In Sec. \ref{sec:vhe} we obtain all the relevant 
magnetic constants for the CrI$_3$ subsystem based on the previously derived magnetic Hamiltonian.  The magnon valley Hall effect and its effects on the transverse
thermal conductivity (via magnon thermal Hall effect) is computed. We
conclude in Sec. \ref{sec:conclusions} with a discussion of our
results.

\section{ Numerical Methods}
\label{sec:methods}

In this section we explained our approach for DFT calculations. We use
the VASP package\cite{vasp1,vasp2,vasp3,vasp4}. The kinetic energy
cutoff is set to 250 eV. The k-points grid is $11\times 11$. For the
exchange-correlation term, we use the SCAN meta-GGA\cite{SCAN}
(actually SCAN+rVV10, see below). At least 12~\AA~ of empty vacuum
space is added to avoid spurious self-interactions along the
non-periodic direction. To account for the Coulomb repulsion of Cr
d-electrons, we use the DFT+U formulation\cite{Liechtenstein95} with
parameters $U=2.7$~eV and $J=0.6$~eV\cite{Lado17}; nevertheless, the
value of magnetic constants is practically unaffected by these
parameters (as long as the system remains insulating). The PyProcar code is employed for the analysis of
eigenvalues.\cite{pyprocar}

To calculate the magnetocrystalline anisotropies, we use projector
augmented-wave (PAW) pseudopotentials \cite{paw}. The spin-orbit
coupling is included in all calculations, including relaxations. The
calculation of the individual anisotropy of each sublattice needs a
non-collinear orientation of the magnetic moments around each Cr atom
in order to explicty break the sublattice symmetry (and hence have
sublattice split anisotropies). This is possible by adding an extra
penalty to the energy:
\begin{equation}
  \Delta E_\epsilon = \epsilon\left[\left[ \bm{m}_a -
    \hat{z}(\bm{m}_a\cdot\hat{z})\right]^2 + \left[ \bm{m}_b -
    \hat{x}(\bm{m}_b\cdot\hat{x})\right]^2\right],\label{eq:mae}
\end{equation}
\noindent with $\bm{m}_\alpha$ the magnetic moment around a Cr ion of
the sublattice $\alpha=\{a,b\}$, \textit{i.e.} $\bm{m}_\alpha =
\int_{R_\alpha} \bm{m}(\bm{r})dr^3$, for some suitable radius
$R_\alpha$ . The $\hat{x}, \hat{z}$ axis refers to in-plane and
out-of-plane directions. The paramenter $\epsilon$ is a factor scaling
the strength of the penalty. The energy from calculations with
different values of $\epsilon$ cannot be directly compared. However,
reversing the sublattice index in Eq. (\ref{eq:mae}), provides a
compatible equation to find the difference between the anisotropies of
both sublattices. The penalty in the energy due to $\epsilon$, can be
decreased (in successive calculations) until it becomes much smaller
than the anisotropy energies. Since the magnetic coupling of Cr ions
is weak, the convergence of the magnetocrystalline energy with
$\epsilon$ was almost immediate.

In a vdW system, it is crucial to have a good description of the
dispersive forces. The SCAN+rVV10 approach\cite{Peng16}, coupling the
SCAN meta-GGA to the revised Vydrov and van Voorhis energy
functional\cite{Sabatini13,Vydrov10} gives an accurate description of
binding energies and lattice parameters of vdW
materials\cite{Tawfik18}.  Our criterion for stopping the structural
relaxation is $0.01$ eV/\AA~ as the largest force. Our electronic
convergence criterion is $10^{-10}$ eV.

\section{\label{sec:model} vdW Heterostructures and Spin Hamiltonian }

In this section we review the CrI$_3$ lattice, we present the geometry
of the proposed heterostructures, and how this translates in a
Heisenberg-like Hamiltonian. The driving force behind the
MVHE is the change of symmetry at the atomic
level due to the heterostructure.

\begin{figure}[h]
\includegraphics[width=0.7\columnwidth]{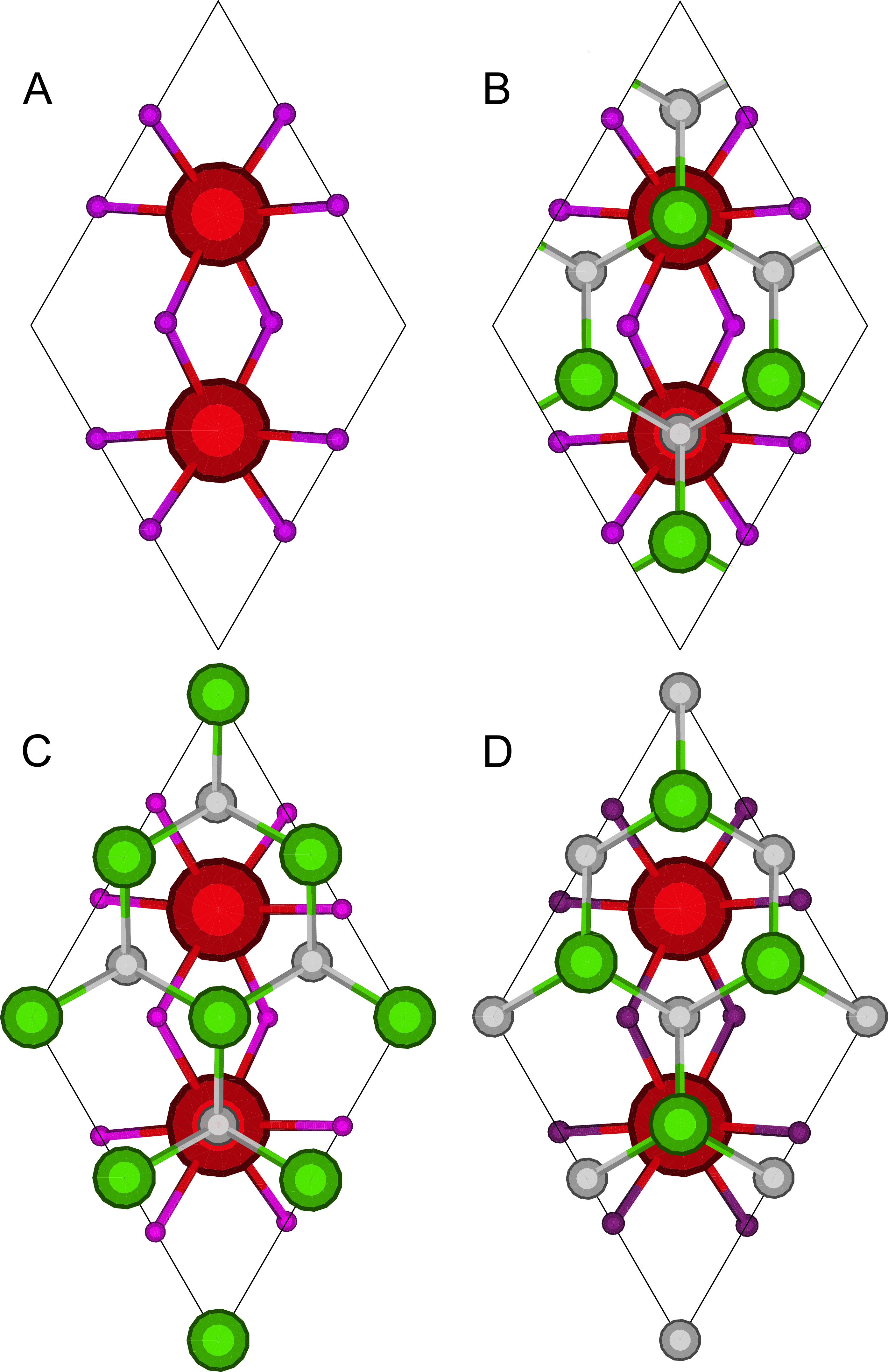}
\caption{Top view of the unit cell of (A) CrI$_3$, Cr(I) atoms are the big red (small purple) spheres. (B-D) Different lateral arrangements of a heterostructure formed by a hexagonal lattice, on top of CrI$_3$. The medium-sized atoms belongs to different
  sublattices of the hexagonal system, white: metal ({e.g.}
  Mo, W, Hf), green: chalcogen ({e.g.} Te, Se, S).}
\label{fig:str}
\end{figure}
CrI$_3$ has a honeycomb lattice, hence two identical sublattices
formed by the Cr ions. Each Cr ion is surrounded by octahedron formed
by six iodine atoms, see Fig.~\ref{fig:str}A. For the magnetic
description (Heisenberg Hamiltonian) the iodine atoms can be ignored,
they are non-magnetic. However, the consequences of altering the
iodine octahedron will be present in the magnetic parameters. The lattice parameter of CrI$_3$ is $a= 6.95$~\AA. If it forms a heterostructure with a two-dimensional (2D) hexagonal system with a
lattice parameter ${a}/{2}$, there is only one arrangement that
preserves the sublattice symmetry of CrI$_3$. However, most hexagonal
2D materials have at least two \textit{different} sublattices ({e.g.} with different elements), and it
is impossible to preserve the sublattice symmetry of CrI$_3$ in the
composed system. The effects of this mechanism on the electronic
structure of CrI$_3$ are small, it is a vdW interaction, but
noticeable on their magnetic properties.

Among the materials whose lattice parameter is nearly ${a}/{2}$ we
can name several transition metal dichalcogenides, with unit formula MX$_2$ (M is a transition metal and X a chalcogen atom) such as: MoTe$_2$ (3.5~\AA), HfS$_2$ (3.6~\AA), TiSe$_2$ (3.5~\AA), WSe$_2$ (3.3~\AA),
WTe$_2$ (3.5~\AA), etc. Other 2D materials also fit: GaS (3.6~\AA), SnS$_2$ (3.6~\AA). It may seem surprising that several 2D materials
have a lattice parameter that is almost ${a}/{2}$. However, the bonding distance in several dichalcogenides ranges from 2.7 to 2.8~\AA,
which coincides with the Cr-I bonding distance, 2.75~\AA. Another way
to understand why the lattice parameter of CrI$_3$ practically doubles
the value found in several hexagonal 2D materials is to consider a
minimal hexagonal and a honeycomb lattices, ignoring all the atomic
detail and only keeping the nodes. If the nodes in both lattices are
at the same distance, the lattice parameter of the honeycomb is the
double of the hexagonal lattice.

\begin{table}[h!]
	\centering
	\caption{Relative binding energy, in meV/unit cell, of the different
		MX$_2$ systems on top of CrI$_3$, see Fig.~\ref{fig:str}. The lowest
		energy arrangement for each system is taken as reference
		(\textit{i.e.} 0 eV).}
	\label{table:de}
	\begin{tabularx}{\columnwidth}{ l rrrrrrrrrrrrrrrrrrrrrrrrrrrrrr } 
		\hline\hline
		&&&& Arrangement &&&&&&&&&&&&&&&&& B &&& C &&&& D && \\ [0.5ex]
		\hline
		&&&& CrI$_3|$MoTe$_2$ &&&&&&&&&&&&&&&&&         4.8  &&&  43.5 &&&& \textbf{0.0} &&\\ 
		&&&& CrI$_3|$HfS$_2$  &&&&&&&&&&&&&&&&&     18.0 &&& 108.9 &&&& \textbf{0.0} &&\\
		&&&& CrI$_3|$WSe$_2$  &&&&&&&&&&&&&&&&&   19.9 &&&  65.8 &&&& \textbf{0.0} &&\\
		\hline
	\end{tabularx}
\end{table}

In the following, we restrict our study to heterostructures of CrI$_3$
over MoTe$_2$, HfS$_2$ and WSe$_2$, in the three arrangements shown in
Figs.~\ref{fig:str}B-D. Even though these materials have the same lattice
and have a very similar composition (metal dichalcogenide), they
produce different effects in the magnetic properties of CrI$_3$. The relative 
 energies among  these arrangements, i.e., taking the lowest energy conformation as reference, are given in
Table~\ref{table:de}. In general, the most stable position is when a chalcogen atom is on top of a Cr atom, and the metals are in bridge positions (arrangement D). The spin Hamiltonian of these heterostructures is a slight variation
of the one proposed by Lado and Fern\'andez-Rossier \cite{Lado17}:
\begin{equation}
H = -\sum_{i\alpha} D_\alpha(S_{i\alpha}^z)^2 -
\sum_{\langle i\alpha,j\beta \rangle}
\left[\frac{J}{2}\vec{S}_{i\alpha}\cdot\vec{S}_{j\beta} +
\frac{\lambda}{2} S_{i\alpha}^zS_{j\beta}^z\right],
\label{eq:hmag}
\end{equation}
where $\langle,\rangle$ stand for summation over next-nearest neighbors lattice sites. The indexes $i,j$ run over each unit cell and $\alpha, \beta$ run over the sublattices $\{a,b\}$.  The value of the spin,
$|\vec{S}_{i\alpha}|$, is $3/2$ ($3\mu_B$ per Cr atom). The first term is the magnetocrystalline anisotropy energy, with $D_\alpha$ depending only on the sublattice, a positive value of $D_\alpha$
implies an out-plane groundstate. The second term is a Heisenberg Hamiltonian, with $J$ being the exchange constant ($J>0$ for ferromagnetic interactions) and $\lambda$ is the exchange anisotropy. According
to Lado\cite{Lado17}, $\lambda$ is the main responsible for the magnetic order of CrI$_3$. To find $J,\lambda$, $D_a$ and $D_b$, we need to evaluate the energy of the FM and antiferromagnetic orders, oriented
in-plane and out-plane. To get the actual value of $D_a$ and $D_b$, we need to explicitly break the sublattice symmetry, by orientating
one sublattice in-plane and the other out-plane (see
Sec.~\ref{sec:methods}). Recently, other spin Hamiltonians have been proposed 
for modelling CrI$_3$\cite{kashin2019,Lee2019}, but the one we are using is 
particularly useful for DFT parameterization and to derive a low-energy magnon Hamiltonian.

The distortion due to the heterostructure also can induce different Cr-I-Cr paths, which in turn can induce a Dzyaloshinskii-Moriya coupling. We calculated the DM vector following the scheme used by Liu \textit{et al.}\cite{Liu2018} for the
groundstate of the CrI$_3|$MoTe$_2$ heterostructure. As the geometry is similar for the other 2D materials, the order of magnitude found also should be similar. The values of $D_z$, the $z$-component of the DM vector, obtained in this way are smaller than $0.01$ meV. Therefore, we will discard the contribution of the DMI to the magnetic Hamiltonian. Even in schemes specifically targeted to enhance DMI, its magnitude is often very small\cite{Ghosh2019}. Another source of DMI comes from the contributions of higher-order neighbors, which should be quite small due to the large interactomic distances.  In bulk CrI$_3$, with magnetic
atoms breaking the inversion symmetry at neighbor layer, DMI can be large enough to be experimentally observed\cite{PhysRevX.8.041028}.

\begin{table}[h!]
	\centering
	\caption{Magnetic constants of the different MX$_2$ systems on top of
		CrI$_3$, in its groundstate arrangement. All the values are in meV,
		see Eq.~\ref{eq:hmag}}
	\label{table:magc}
	\begin{tabularx}{\columnwidth}{ l  rrrrrrrrrrrrrrrrrrrrrrrrrrrr} 
		\hline\hline
		&&&& System &&&&&&&&&&&& J &&& $\lambda$ &&& D$_a$ &&& D$_b$ &&&\\ 
		\hline
		&&&& CrI$_3$          &&&&&&&&&&&& 2.20  &&&  0.11 &&&  0.04 &&&  0.04 &&&\\
		&&&& CrI$_3|$MoTe$_2$ &&&&&&&&&&&& 2.43  &&&  0.04 &&& -0.18 &&&  0.33 &&&\\ 
		&&&& CrI$_3|$HfS$_2$  &&&&&&&&&&&& 2.05  &&&  0.13 &&& -0.04 &&& -0.07 &&&\\
		&&&& CrI$_3|$WSe$_2$  &&&&&&&&&&&& 2.32  &&&  0.04 &&&  0.04 &&&  0.07 &&&\\
		\hline
	\end{tabularx}
\end{table}

The values of the magnetic constants, in the groundstate (atomic)
configuration, are shown in Table \ref{table:magc}. The value of $J$
 varies only in about 10 \%, since no fundamental change in the
electronic structure happens. The anisotropy $\lambda$ is much more
affected, decreasing to half its value in CrI$_3|$MoTe$_2$ and
CrI$_3|$WSe$_2$. This is explained by the distortion on CrI$_3$ once in contact with another material, for instance, both Cr sublattices
no longer are co-planar. Nevertheless, the value of $\lambda$ remains positive, 
indicating a FM easy axis. Finally, the magnetocrystalline anisotropy
suffers strong variations, especially in MoTe$_2$. The strong variations in the anisotropy constants can be expected: alterations of the symmetry of the local environment (even in vdW systems) can induce large changes, even inducing an easy axis\cite{munoz10}. In some of the heterostructures studied, the
change of $D_\alpha$ also includes a change of its sign ($D_\alpha <
0$), this does not means a change of easy axis, since the exchange anisotropy $\lambda$ dominates and is positive in every heterostructure. 

\section{\label{sec:vhe} Magnon Valley Hall Effect }

In this section we study spin fluctuations in the limit of small
deviations (magnons) around the equilibrium state.  We consider the
ground state to be collinear and parallel to the
$z$-direction. Magnonic excitations are introduced by the standard
Holstein-Primakoff \cite{HolsteinPR1940} transformation that quantizes
the spins in terms of bosons\cite{HPmagnons}. 

\begin{figure}[h!]
\includegraphics[width=1\columnwidth]{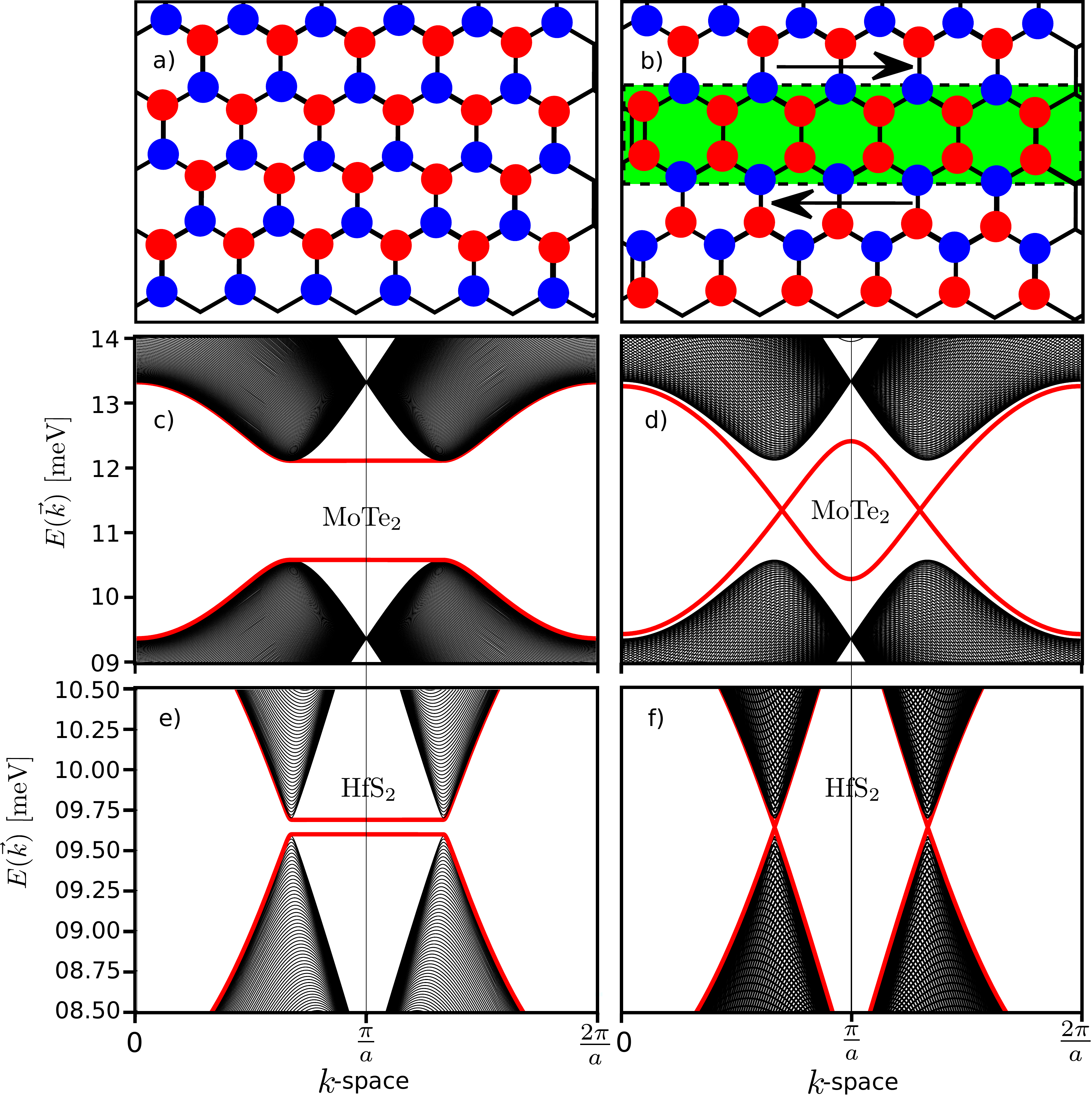}
\caption{(a) Bulk honeycomb lattice formed by the Cr atoms, different
  onsite terms in Eq.~\ref{eq:sham} are marked by different
  colors. (b) A 1D domain wall swapping the sublattices
  (\textit{i.e.}, the onsite term in the spin Hamiltonian), the
  topologically protected magnon states will appear around this
  region. Magnon edge states (red lines) and subbands (black lines) of
  a finite ribbon of CrI$_3|$MoTe$_2$ and CrI$_3|$HfS$_2$ without (c,
  e) and with (d, f) a domain wall. With the domain wall, the
  topological edge states bridge the gap in both cases.}
\label{fig:Valley}
\end{figure}

The spin Hamiltonian given by Eq. (\ref{eq:hmag}) can be expanded up to second order in magnon operators resulting in ${H}=H_0+H_m$. The zero-point
energy is represented by $H_0$, while the nearest neighbor tight-binding Hamiltonian
for the magnonic excitations is $H_m=-\frac{JS}{2}\sum_{\langle i,j\rangle}\left(d_i^\dagger d_j+h.c\right)+\Omega\sum_i d_i^\dagger d_i$, in the absence of substrates and where $\Omega=S(2D+3(J+\lambda))$. The magnon
operator $d_{i}(d_{i}^\dagger)$ corresponds to the annihilation
(creation) operator at  the $i$-th site. The Hamiltonian
$H_m$ is similar to the electronic Hamiltonian of
graphene, with two Dirac points existing at ${\bf
  K}^+=\left(2\pi/\sqrt{3}a,0\right)$ and ${\bf
  K}^-=\left(-2\pi/\sqrt{3}a,0\right)$ in the Brillouin zone. In CrI$_3$, a magnon gap $\Delta_0=3S\lambda= 0.4$ meV for the lower 
energy band and critical temperature of $T_c=85$K are
found\cite{Lado17}. Under the presence of the proposed substrates, the Cr environment changes within the unit cell, thus modifying the magnetocrystalline anisotropy energy in each sublattice. The effect on
the Hamiltonian $H_m$ 
is readily captured by mapping the magnon operators into the sublattice-magnon basis,
\begin{equation}
H_m=-  JS\sum_{\langle i,j\rangle}\left(a_i^\dagger b_j+h.c \right)+\Omega_a\sum_i a_i^\dagger a_i+ \Omega_b\sum_i  b_i^\dagger b_i,
\label{eq:sham}
\end{equation}
where $a^\dagger$ and $b^\dagger$ represent magnon creation operators on the
sublattices $a$ and $b$, respectively. Also, we defined
$\Omega_{\alpha}=\left[2D_{\alpha}S+3S(J+\lambda)\right]$. The induced
effect by substrates on the magnon bands can be readily captured in
the momentum representation. In Fourier space the magnon Hamiltonian
then reads $H_m=\sum_{\bm{k}}\Psi^{\dagger}_{\bm{
  k}}\left[\bar{\Delta}\mathbb{I}+{\bm{h}}_{\bm{
    k}}\cdot{\boldsymbol{\tau} }\right]\Psi_{\bm{ k}}$, with $\Psi_{\bm{
  k}}=(a_{\bm{k}},b_{\bm{k}})$ the spinor of Fourier transformed
operators and $\boldsymbol{\tau}$ the Pauli matrix vector. The field
${\bm{h}}_{\bm{ k}}=\sum_j\left(-JS\cos({\bm{k}}\cdot{\bm{
  v}}_j),JS\sin({\bm{k}}\cdot{\bm{u}}_j),\Delta\right)$, with
$\bar{\Delta}=\left(\Omega_a+\Omega_b\right)/2$ and
${\Delta}=\left(\Omega_a-\Omega_b\right)/2$. The eigenenergies for the
upper and lower magnon bands are given by $\epsilon^{\pm}_{\bm{k}}=\bar{\Delta}\pm \left|{\bm{h}}_{\bm{k}}\right|$. In the limit
$\Omega_a=\Omega_b$ both bands become degenerated at the Dirac points ${\bf K}^+$ and ${\bf K}^-$, with a Dirac-type dispersion around these points.  In the presence of the substrates, the magnetic anisotropy
becomes different on each sublattice and a band-gap is open at ${\bf K}^+({\bf K}^-)$ with value $2\Delta$. It is worth noticing that the gaps open differently since $\epsilon^{{\bf K}^+}_{\pm}=\bar{\Delta}\pm \Delta$ and $\epsilon^{{\bf K}^-}_{\pm}=\bar{\Delta}\mp \Delta$.  The Berry curvatures of the upper and the lower
magnon bands are largely concentrated, and opposites in sign, around the corners of the BZ\cite{Berry}. Thus, the Chern number of each band is zero\cite{Wang2018b}. However, restricting the integration zone to a single valley, around ${\bf K}^+$, the Chern number is $c_n^\pm = \int_{\text{}} \Omega_z^\pm dk^2 = \pm 1$\cite{Berry}. 
  
Although perpendicularly magnetized honeycomb lattices with staggered anisotropy on each sublattices are topologically trivial, this represents the basic ingredient to induce a {magnon valley Hall effect} in a CrI$_3$-based vdW heterostructure. Motivated by related works on graphene-like structures \cite{Yao2009}, we consider an induced \textit{band inversion} as a second ingredient. This consists in a sign change of the bandgap obtained, e.g., swapping the sublattices $a\leftrightarrow b$, that in turn
implies swapping the anisotropy $D_a\leftrightarrow D_b$. The last can be
achieved by inducing a line defect on the MX$_2$ monolayer, see
Fig.~\ref{fig:Valley}b. This type of defect has been extensively studied, being the most common methods to induce this inversion those based on chalcogen defects \cite{Alvarez18,MArika18,Komsa13}, see Ref. [\onlinecite{Batzill18}] and references therein. Recently,  a sublattice inversion induced by irradiation has been reported for MoTe$_2$.\cite{Elibol18}. Another method to create a post-synthesis sublattice inversion is by means of the incorporation of excess Mo atoms in MoTe$_2$. This induces self-organization into highly-ordered 1D patterns that reach a length of several nanometers.\cite{Coelho18}

In Figs. \ref{fig:Valley}c and \ref{fig:Valley}e we show, in a zigzag nanoribbon without line defect, that the edges states are flat bands connecting the $\bf{K}^+$ and $\bf{K}^-$ valleys. The bandgap achieved by this vdW heterostructure goes from 0.1 meV  (CrI$_3|$HfS$_2$) to nearly 1.5 meV (CrI$_3|$MoTe$_2$). Depending on the actual nanoribbon termination, extra edges states can appear (due to non-bonding atoms in
the chemical jargon)\cite{Li2012}.
\begin{figure}[h]
\includegraphics[width=1\columnwidth]{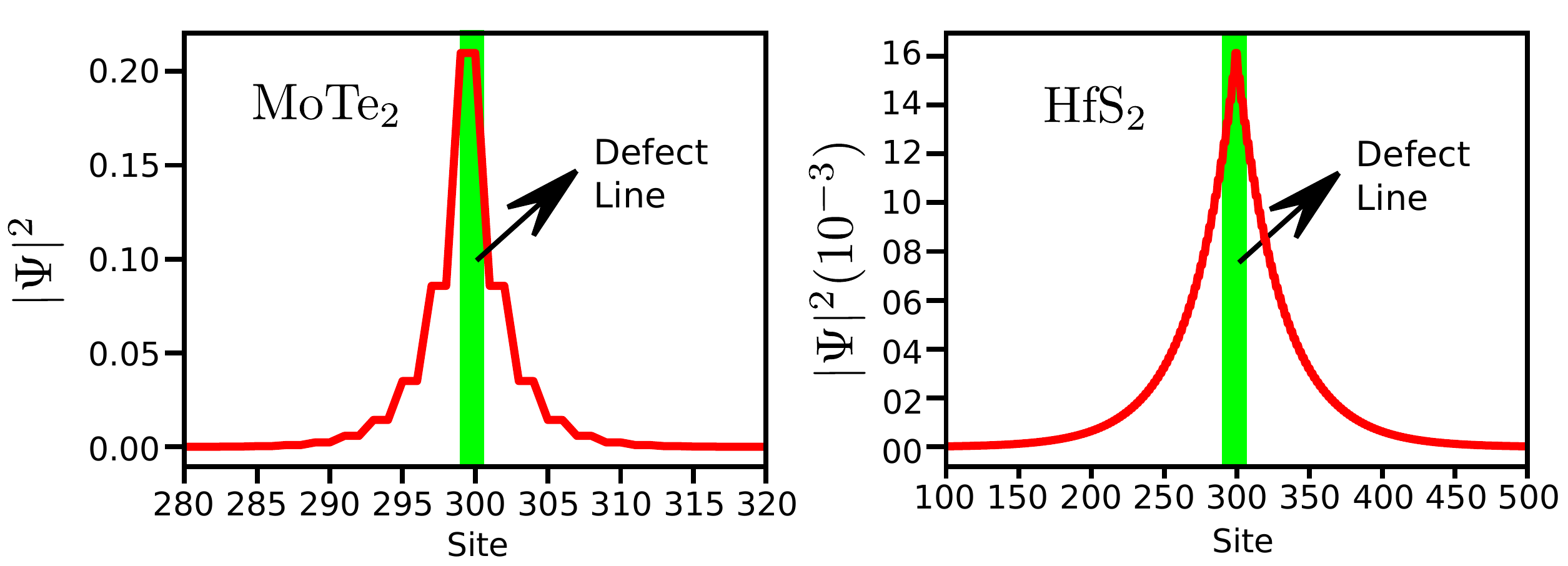}
\caption{ Topological protected states wavefunction at defect line near proyection of ${\bf K}^+$ point, left panel show CrI$_3|$MoTe$_2$ wavefunction (large band-gap) and right panel CrI$_3|$HfS$_2$ wavefunction (small band-gap). The localization length is much larger for systems with smaller magnon bandgap.}
\label{fig:wave}
\end{figure}

When sublattice inversion is considered by introducing a line defect (Fig. \ref{fig:Valley}b)  and the condition ${J}/{4} \geq |D_a-D_b|$ is fulfilled, the magnon valley hall effect (MVHE) is presented in the system. Two topologically protected states with opposite velocities at the defect line appear connecting the valence and conduction bands (Fig. \ref{fig:Valley}d,f). Particularly, CrI$_3$ with all substrates meet the necessary condition to exhibit this effect. These states at the defect line (see Fig. \ref{fig:wave}) have a exponential decay that depends on the band-gap induced by the substrate without defect line. While the topological states are localized, its penetration depth is strongly dependent of the topological bandgap. The localization of these states have deep consequences in presence of disorder braking the sublattice symmetry (\textit{i.e.} disorder in anisotropies). When the edge state penetrates several sites, the atomic disorder averages to zero, making the edge state mostly unaffected even in the case oa disorder-induced bandgap, even for very large vlaues of the disorder\cite{munoz2018topological}.  

In order to connect the MVHE with experimentally accessible measurements, we consider magnon transport in the vdW heterostructures in presence of a thermal bias. In linear response, we compute the magnon thermal Hall conductivity, $\kappa_{xy}$\cite{Laurell18}, under a longitudinal temperature gradient. Following standard transport theory, we compute the thermal conductivity given by the expression, $  \kappa_{xy}=-{k_B^2 T/h^2}\sum_{n=\pm}\int\left(c_2(g(\epsilon_{nk}))-{\pi^2}/{3}\right)\Omega^n_z(k)d\bm{k}$, where the sum runs over both eigenvalues, $\epsilon_\pm$, and $g(\epsilon)$ is the Bose-Einstein distribution. The function $c_2$\cite{c2function} is a monotonous function satisfying $c_2(0)\to 0$ and $c_2(\infty)\to {\pi^2}/{3}$. The value of $\kappa_{xy}$ integrated over all the Brillouin zone is exactly zero, since both valleys have an opposite Berry curvature. If we restrict the integral to a neighborhood around ${\bf K}^+$, we will get the contribution of each valley to the Hall thermal conductivity. The valley effect of magnons arises when a defect line swaps both sublattices, following the mechanism introduced above. 
\begin{figure}
    \centering
    \includegraphics[width=\columnwidth]{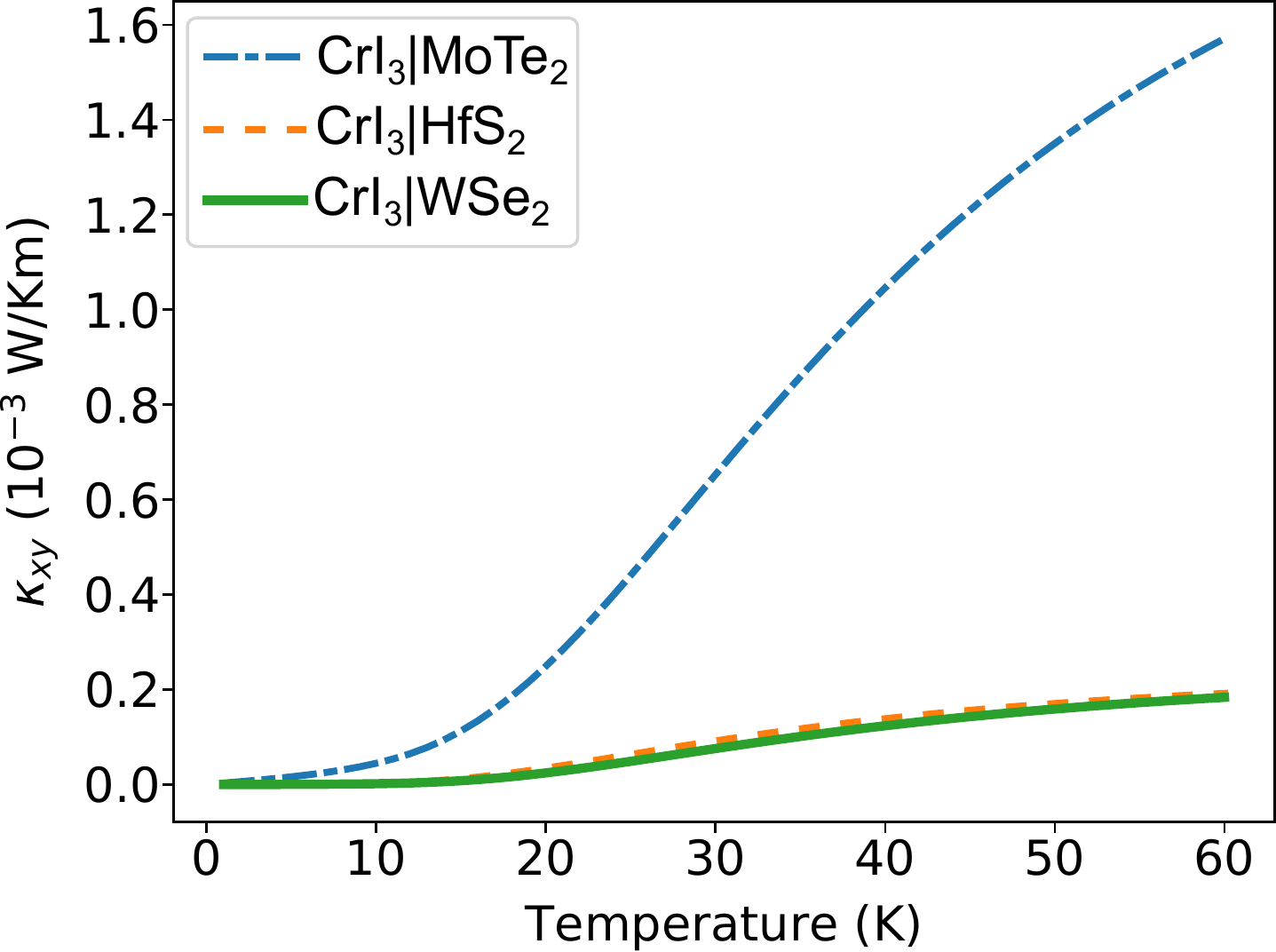}
    \caption{Temperature-dependent magnon thermal Hall conductivity $\kappa_{xy}$ for each of the proposed vdW heterostructures. A thickness of 1 nm was used to normalizes $\kappa_{xy}$. The temperature range is intentionally considered up to the critical temperature $T_C=85$K.}
    \label{fig:thermal}
\end{figure}
In Fig. \ref{fig:thermal} we shows the temperature-dependent thermal conductivity. 
The result of $\kappa_{xy}$ for every vdW heterostructure is in the range of $10^{-3}$ (W/Km), close to the thermal conductivity of other topological schemes\cite{Laurell18}. Importantly, when the temperature increases we see a significant upturn of $\kappa_{xy}$ for CrI$_3|$MoTe$_2$, contrary to the other heterostructures due to their small gap. As is expected, the thermal conductivity reaches a maximum value, roughly at a similar critical temperature, then decay to zero for larger temperature.

\section{\label{sec:conclusions}Conclusions}
In this paper, we proposed a mechanism to induce topologically non-trivial states in the magnon spectrum of a single layer of CrI$_3$, in analogy with the valley Hall effect in the electronic structure of bilayer graphene. The topological edge states are achieved when \textit{(i)} each sublattice has a different magnetocrystalline anisotropy and \textit{(ii)} there exists a region where the magnetocrystalline of both sublattices swaps. A sublattice-dependent magnetocrystalline anisotropy, is obtained by forming a vdW heterostructure with another (non-magnetic) hexagonal 2D material or substrate. The local sublattice inversion requires a line defect in the CrI$_3$ layer. These defects, with a extension of several nanometers, can be found naturally or artificially induced with great accuracy. We quantified this effect by DFT calculations of the heterostructures of CrI$_3$ with MoTe$_2$, HfS$_2$ and WSe$_2$. The topological bandgap induced in the magnon spectrum in the previous materials ranged between 0.1 to 1.5 meV.
Finally, we show that the MVHE manifests itself in the form of a magnon thermal Hall effect, due to the non-trivial topology of the band structure, when a thermal gradient is applied along the heterostructure. This effect is quantified by the evaluation of the temperature-dependent transverse thermal conductivity.


\begin{acknowledgments}

This work was partially funded by Fondecyt grants 1190036 (EEV),
1191353 (FM), 11180557 (RIG), Conicyt doctoral fellowship grants 21151207 (JM), the Center for the Development of Nanoscience and
Nanotechnology CEDENNA FB-0807, the supercomputing infrastructure
of the NLHPC (ECM-02) and from Conicyt PIA/Anillo ACT192023 (FM).
R.E.T acknowledges the support by the European
Union's Horizon 2020 Research and Innovation Programme under Grant
DLV-737038 "TRANSPIRE" and the Research Council of Norway through is
Centres of Excellence funding scheme, Project No. 262633,
"QuSpin". The authors thank fruitful discussion with Luis E. F. Foa Torres.

\end{acknowledgments}

\bibliography{bib}

\begin{thebibliography}{60}
\bibitem [{HPm()}]{HPmagnons}%
  \BibitemOpen
  \href@noop {} {\bibinfo  {journal} {The Holstein-Primakoff transformation is
  defined by ${S}^+_{i}=\sqrt{2S-{{d}^{\dagger}_{i}{d}_{i}}} {d}_{i}$, $
  {S}^-_{i}={d}^{\dagger}_{i}\sqrt{2S-{{d}^{\dagger}_{i}{d}_{i}}}$ and
  ${S}^z_{i}=\left(S-{ d}^{\dagger}_{{i}}{d}_{{i}}\right)$ with $S$ the spin
  quantum number. One magnon, created (annihilated) at site ${i}$ of the
  lattice by the operator ${ d}^{\dagger}_{{i}}$(${ d}_{{i}}$), corresponds to
  changing the total spin with $+\hbar$($-\hbar$)}\ }\BibitemShut {NoStop}%
\bibitem [{c2f()}]{c2function}%
  \BibitemOpen
  \href@noop {} {\bibinfo  {journal} {The function $c_2$ is defined as,
  $c_2(x)=(1+x)\left[\text{ln}\left(\frac{1+x}{x}\right)\right]^2-\left(\text{ln}(x)\right)^2
  -2\text{Li}_2(-x)$, where $\text{Li}_n$ is the polylogarithm function}\
  }\BibitemShut {NoStop}%
\end{thebibliography}

\end{document}